\documentclass[epj]{svjour}
\usepackage{graphics}
\usepackage{amssymb}

\begin{document}
\newcommand{\bvec}[1]{\mbox{\boldmath ${#1}$}}
\title{Multipole approach for photo- and electroproduction of kaon}

\author{T. Mart\inst{1} \and A. Sulaksono\inst{1}}
\institute{Departemen Fisika, FMIPA, Universitas Indonesia, Depok 16424, 
  Indonesia}
\date{Received: date / Revised version: date}
\abstract{
  We have analyzed the experimental data on $K^{+}\Lambda$ 
  photoproduction by using a multipole approach. In this analysis we use the 
  background amplitudes constructed from appropriate Feynman diagrams in a 
  gauge-invariant and crossing-symmetric fashion. Results of our analysis 
  reveal the problem of mutual consistency between the new {\small SAPHIR} and {\small CLAS} 
  data. We found that the problem could
  lead to different conclusions on ``missing resonances''. We have also
  extended our analysis to the finite $Q^2$ region and compared the result
  with the corresponding electroproduction data.
   % end of PACS codes
} %end of abstract
\maketitle
\section{Introduction}
\label{intro}
In the last decades, there have been a large number of attempts devoted to
understand hadronic interactions in the medium energy region. However,
due to the nonperturbative nature of QCD at these energies, hadronic physics
continues to be a challenging field of investigation.

One of the most intensively studied topics in the realm of hadronic
physics is the associated strangeness photoproduction. High-intensity 
continuous electron beams produced by modern accelerator technologies, 
along with unprecedented precise detectors, are among the important
aspects that have brought renewed attention to this subject. 

On the other hand, the argument that some of 
the resonances predicted by constituent quark
models are strongly coupled to strangeness channels, and therefore
intangible to $\pi N \to \pi N$ reactions that are used by Particle
Data Group ({\small PDG}) to extract the properties of nucleon resonances, has raised
the issue of ``missing'' resonances. As a consequence, recent analyses of 
strangeness photoproduction have mostly focused on the quest
of missing resonances \cite{Mart:1999ed}.
With the new {\small CLAS} data appearing this year
\cite{Bradford:2005pt}, this becomes an arduous task, since 
several recent phenomenological studies found a lack of mutual 
consistency between the recent {\small CLAS} and {\small SAPHIR} \cite{Glander:2003jw} data.

In view of this, it is certainly important to investigate the 
physics consequence of using each data set. Ideally, this should 
be performed on the basis of a coupled-channels formalism. However, 
the level of complexity in such a framework increases quickly 
with the addition of resonance states. It is widely known that
in kaon photoproduction too many resonance states can contribute,
whereas there is a lack of systematic procedure to determine how many
resonances should be built into the process. Thus, for this purpose, we 
constrain the present work to a single-channel analysis, but we use 
as much as possible nucleon resonances listed by {\small PDG}. This has the 
advantage that we can simultaneously explore the importance 
of higher spin states in kaon photoproduction.

\section{Kaon Photoproduction}
\label{sec:Kaon_Photoproduction}
\subsection{Formalism}
\label{subsec:Formalism}
The background amplitudes are obtained from a series of tree-level Feynman 
diagrams. They consist of the standard $s$-, $u$-, and $t$-channel Born terms 
along with the $K^*(892)$ and $K_1(1270)$ $t$-channel vector mesons. 
Altogether they are often called extended Born terms.

The resonant multipoles for a state with the mass $M_R$, width 
$\Gamma$, and angular momentum $\ell$ are assumed to have the Breit-Wigner form
\cite{Tiator:2003uu}
\begin{eqnarray}
  \label{eq:em_multipole}
  A_{\ell\pm}^R(W) = {\bar A}_{\ell\pm}^R  c_{KY} \frac{f_{\gamma R}(W)
    \Gamma_{\rm tot}(W) M_R\, f_{K R}(W)}{M_R^2-W^2-iM_R\Gamma_{\rm tot}(W)} e^{i\phi},~ 
  \label{eq:m_multipole}
\end{eqnarray}
where $W$ represents the total c.m. energy, the isospin factor 
$c_{KY}$ is $-1$, $f_{KR}$ is the usual
Breit-Wigner factor describing the decay of a resonance $R$ with a total width
$\Gamma_{\rm tot}(W)$ and physical mass $M_R$. The $f_{\gamma R}$ indicates
the $\gamma NR$ vertex and $\phi$ represents the phase angle. 
The Breit-Wigner factor $f_{KR}$ is given by 
\begin{eqnarray}
  \label{eq:f_KR}
  f_{KR}(W) &=& \left[\frac{1}{(2j+1)\pi}\frac{k_W}{|\bvec{q}|}\frac{m_N}{W}
  \frac{\Gamma_{KY}}{\Gamma_{\rm tot}^2}\right]^{1/2}~,
\end{eqnarray}
with $m_N$ and $k_W$ indicating the nucleon mass and the photon equivalent energy, 
respectively. The energy dependent partial width $\Gamma_{KY}$ is defined through
\begin{eqnarray}
  \label{eq:Gamma_KY}
  \Gamma_{KY} &=& \beta_K\Gamma_R \left(\frac{|\bvec{q}|}{q_R}\right)^{2\ell+1}
  \,\left(\frac{X^2+q_R^2}{X^2+\bvec{q}^2}\right)^\ell\frac{W_R}{W}~,
\end{eqnarray}
where the damping parameter $X$ is assumed to be 500 MeV for all resonances,
$\beta_K$ is the single kaon branching ratio, 
$\Gamma_R$ and $q_R$ are the total width and kaon c.m. momentum at $W=M_R$. 
The $\gamma NR$ vertex is parameterized through
\begin{eqnarray}
  \label{eq:f_gammaR}
  f_{\gamma R} &=& \left(\frac{k_W}{k_R}\right)^{2\ell '+1}\,\left(\frac{X^2+k_R^2}{
    X^2+k_W^2}\right)^{\ell '} ~,
\end{eqnarray}
where $k_R$ is equal to $k_W$ calculated at $W=M_R$. For $M_{\ell\pm}$ and 
$E_{\ell +}$: $\ell' =\ell$, whereas for $E_{\ell -}$: $\ell '= \ell -2$ if 
$\ell \ge 2$. The values of $\ell$ and $\ell '$ as 
well as other parameters are given in Ref.\,\cite{Mart:2006dk}.
All observables can be calculated from the CGLN amplitudes
\begin{eqnarray}
  \label{eq:cgln}
  {\cal F}\!&=&\!\bvec{\sigma} \cdot \bvec{b}F_1 - i \bvec{\sigma} \cdot 
  \hat{\bvec{q}}~\bvec{\sigma} \cdot (\hat{\bvec{k}} \times \bvec{b})F_2
  + \bvec{\sigma} \cdot \hat{\bvec{k}}~\hat{\bvec{q}} \cdot \bvec{b} F_3
 \nonumber\\ && 
 +\bvec{\sigma} \cdot \hat{\bvec{q}}~\hat{\bvec{q}} \cdot \bvec{b} F_4 
  - \bvec{\sigma} \cdot \hat{\bvec{q}}~b_0 F_5 - \bvec{\sigma} \cdot 
   \hat{\bvec{k}}~b_0 F_6 
\end{eqnarray}
where $b_{\mu}=\epsilon_{\mu}-({\hat{\bvec{k}} \cdot 
  \bvec{\epsilon}}/{|\bvec{k}|})k_{\mu}$. For 
photoproduction only the first four 
amplitudes $F_i$ are relevant. They are 
related to the electric and magnetic multipoles
given in Eq.\,(\ref{eq:em_multipole}) by
\begin{eqnarray}
F_1 &=& \sum_{\ell \ge 0} \left\{ \left( \ell M_{\ell+}\!+ E_{\ell+} \right) P'_{\ell\!+\!1}
+ \left[ \left( \ell+1 \right) M_{\ell-}\! +
E_{\ell-} \right] P'_{\ell-1} \right\}\!, \nonumber\\
F_2 &=& \sum_{\ell \ge 1} \left[ \left( \ell + 1 \right) M_{\ell+}
+ \ell M_{\ell-} \right] P'_{\ell} , \nonumber\\[-3ex]\label{eq:cgln1}\\
F_3 &=& \sum_{\ell \ge 1} \left[ \left( E_{\ell+} - M_{\ell+} \right) P''_{\ell+1}
+ \left( E_{\ell-} + M_{\ell-} \right) P''_{\ell-1} \right] , \nonumber\\
F_4 &=& \sum_{\ell \ge 2} \left( M_{\ell+} - E_{\ell+} - M_{\ell-} - E_{\ell-} \right)
P''_{\ell} .\nonumber
\end{eqnarray}

\subsection{Numerical Results}
\label{subsec:Results}
The number of free parameters is relatively large. To reduce this
we fix both $g_{K\Lambda N}$ and $g_{K\Sigma N}$ coupling constants
to the SU(3) predictions and fix masses as well as widths of the
four-star resonances to their {\small PDG} values. Due to the problem of mutual 
consistency between {\small SAPHIR} and {\small CLAS} data, in the fitting procedure we 
define two different data sets. In the first set (Fit 1) we use the {\small SAPHIR} 
and {\small LEPS} data, while in the second one (Fit 2) we use the  
{\small CLAS} and {\small LEPS} data. In total we use 15 nucleon resonances listed by {\small PDG}.
The $\chi^2$ minimization fit is performed by using the {\small CERN-MINUIT} code.

\begin{figure}
\resizebox{0.5\textwidth}{!}{\includegraphics{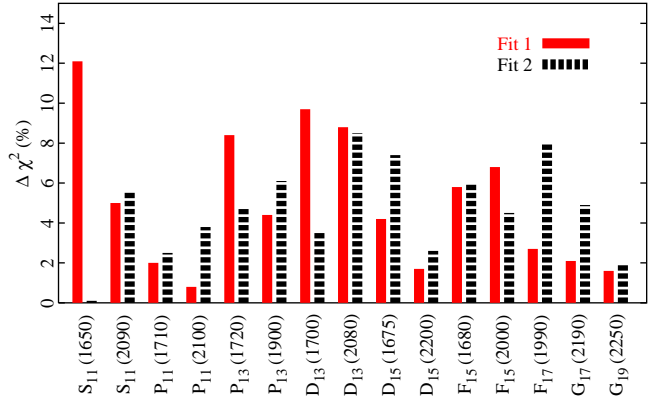}}
\caption{Significance of the individual resonances in our fits.}
\label{fig:strength}
\end{figure}

To further investigate the importance of the individual resonances we define a parameter
\begin{eqnarray}
  \label{eq:par_res}
  \Delta \chi^2 ~=~ \frac{\chi^2_{\rm All}-\chi^2_{{\rm All}-N^*}}{\chi^2_{\rm All}}
  \times 100\,\% ~,
\end{eqnarray}
where $\chi^2_{\rm All}$ is the $\chi^2$ obtained by using all resonances 
and $\chi^2_{{\rm All}-N^*}$ is the $\chi^2$ obtained by using all 
but a specific resonance. Therefore, $\Delta \chi^2$ 
measures the relative difference between the $\chi^2$ of including and of 
excluding the corresponding resonance. The result is shown in the histogram of 
Fig.\,\ref{fig:strength}. 

Except for $S_{11}(2090)$, $P_{11}(1710)$, 
$D_{13}(2080)$, $F_{15}(1680)$, and $G_{19}(2250)$, for which 
the $\Delta\chi^2$ are almost similar, the histogram shows that 
the new {\small CLAS} and {\small SAPHIR} data can be only explained by different sets of 
nucleon resonances. Constraining the $\Delta\chi^2 \gtrsim 6\%$, e.g., 
leads to the fact that the important resonances in Fit 1 are the $S_{11}(1650)$, 
$P_{13}(1720)$, $D_{13}(1700)$, $D_{13}(2080)$, $F_{15}(1680)$, 
and $F_{15}(2000)$, while Fit 2 needs the $P_{13}(1900)$, $D_{13}(2080)$, 
$D_{15}(1675)$, $F_{15}(1680)$, and $F_{17}(1990)$. 

It is interesting to note here
that both Fit 1 and Fit 2 support the requirement of the 
$D_{13}(2080)$ in this process. Surprisingly,
all new data reject the need for the $P_{11}(1710)$, while the new {\small CLAS} data 
do not require the $P_{13}(1720)$ resonance. Although
most recent analyses of the $K^+\Lambda$ channel have included these 
intermediate states, this conclusion corroborates the finding of
Ref.\,\cite{Arndt:2003ga}.

Another new phenomenon is the contribution 
from the $F_{15}(2000)$ and $F_{17}(1900)$, which are quite important 
according to {\small SAPHIR} and {\small CLAS} data, respectively.
These resonances have not been used in most analyses, especially in the isobar
model with diagrammatic technique, since propagators for spins 5/2 and 7/2 
are not only quite complicated in this approach, but also their forms 
are not unique. 

\begin{figure}
\resizebox{0.5\textwidth}{!}{\includegraphics{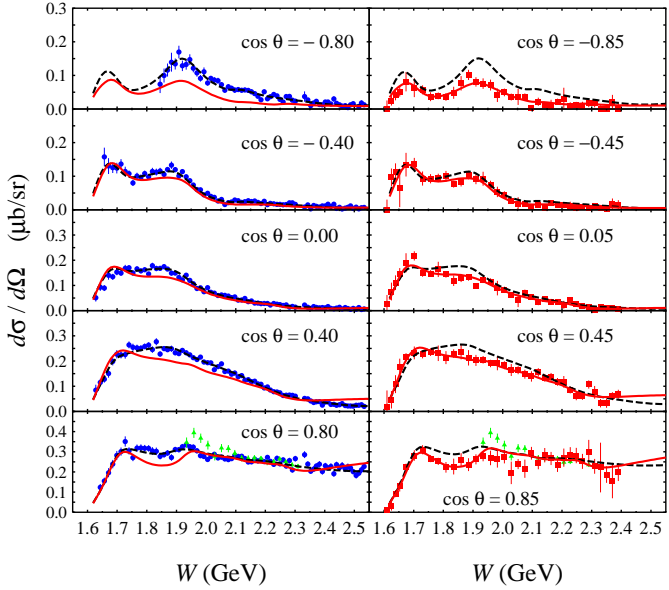}}
\caption{Differential cross sections obtained from Fit 1 (solid curves)
  and Fit 2 (dashed curves) as a function of the total c.m. energy.
  Solid circles, squares, and triangles represent experimental data
  from the {\small CLAS}, {\small SAPHIR}, and {\small LEPS} collaborations, respectively.}
\label{fig:diff}
\end{figure}

\begin{figure}
\resizebox{0.5\textwidth}{!}{\includegraphics{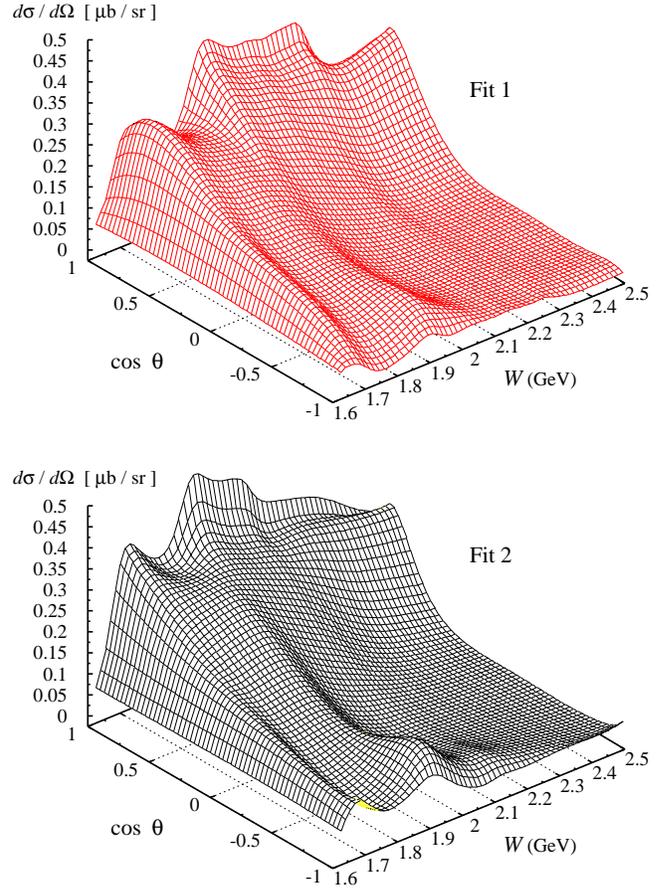}}
\caption{Differential cross sections obtained from Fit 1 and Fit 2
as functions of $W$ and $\cos\theta_K$.}
\label{fig:diff3d}
\end{figure}

In Fig.~\ref{fig:diff} we show the comparison
between the results of Fit 1 and Fit 2 with the {\small CLAS} data. 
The {\small LEPS} data are also shown in this case. It is obvious 
from the figure that the {\small LEPS} data are closer to 
the {\small CLAS} data. From this figure it is also clear that the 
largest discrepancy appears between $W=1.75$ GeV and 1.95 GeV in the forward
direction, whereas in the backward direction the discrepancies show up
in a wider range, i.e., from 1.8 to 2.4 GeV. It is also important to note that
at the very forward and backward angles the two data sets 
exhibit very different trends. The {\small CLAS} data
tend to rise at these regions, while the {\small SAPHIR} 
data tend to decrease. Nevertheless, this does not happen 
in the whole energy region. Especially near the forward angles,
where we found that the result of Fit 2 (the new
{\small CLAS} data) shows more structures than that of Fit 1.
To have a better view of this, in Fig.\,\ref{fig:diff3d} we 
present the three-dimensional plot of 
the differential cross section as 
functions of $W$ and $\cos\theta$. 

\begin{figure}
\begin{center}
\resizebox{0.4\textwidth}{!}{\includegraphics{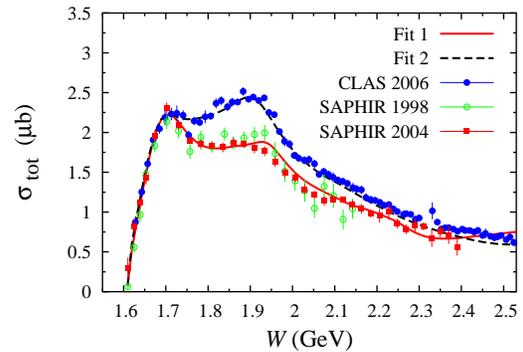}}
\caption{Comparison between experimental total cross sections 
  with the predictions of Fit 1 and Fit 2. Experimental data
  shown in this figure were not used in the two fits.}
\label{fig:total}
\end{center}
\end{figure}

Prediction for the total cross sections of both fits is  
shown in Fig.\,\ref{fig:total}. From this figure it is clear
that the extracted total cross sections from both 
collaborations are consistent with their differential
cross sections. 

We have investigated the origin of the second peak
at $W\approx 1.9$ GeV as shown in Fig.\,\ref{fig:total}. The result
indicates that in both fits the peak originates from the 
$D_{13}(2080)$ with a mass of 1936 MeV if {\small SAPHIR} data were used
or 1915 MeV if {\small CLAS} data were used. This result is in good agreement
with the finding from several recent studies 
\cite{Mart:2004ug,Bartholomy:2004uz,Julia-Diaz:2006is}

\section{Kaon Electroproduction}
\label{sec:Kaon_Electroproduction}
For kaon electroproduction the longitudinal amplitudes
$F_5$ and $F_6$ should be taken into account in Eq.\,(\ref{eq:cgln}).
These amplitudes are given by
\begin{eqnarray}
  \label{eq:f5}
  F_5 &=& \sum_{\ell \ge 0} \left\{ (\ell+1)L_{\ell+}P'_{\ell+1} - \ell L_{\ell-} P'_{\ell-1}\right\}, \\
  \label{eq:f6}
  F_6 &=& \sum_{\ell \ge 1} \left\{ \ell L_{\ell-} - (\ell+1) L_{\ell+}\right\} P'_{\ell},
\end{eqnarray}
where the longitudinal multipoles are related to the scalar ones   
by $L_{\ell\pm}=k_0S_{\ell\pm}/|\bvec{k}|$.

\begin{figure}
\begin{center}
\resizebox{0.45\textwidth}{!}{\includegraphics{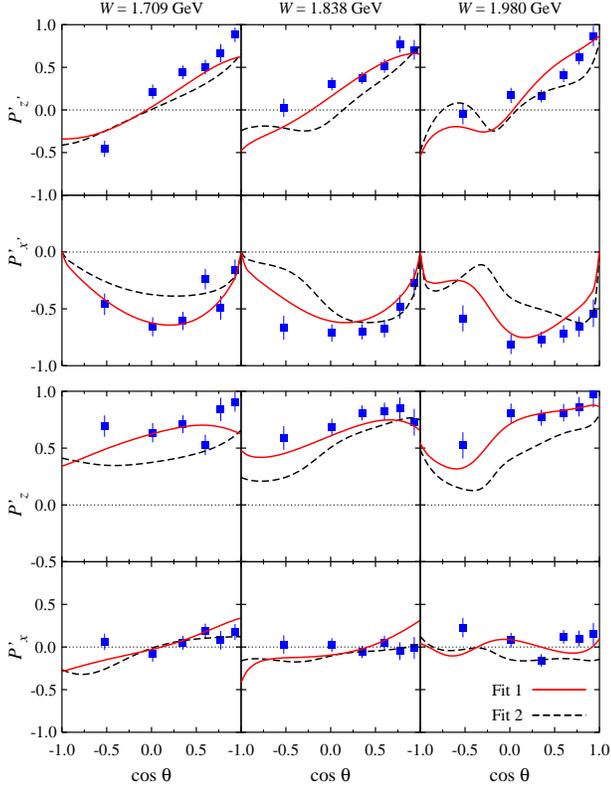}}
\caption{Transfered $\Lambda$ polarization components $P'_{z'}$
  and $P'_{x'}$ (upper panels) and $P'_{z}$ and $P'_{x}$ (lower panels) obtained
  from the two models compared with the new {\small CLAS} measurement \cite{Carman:2002se}.
  Here, both $Q^2$ and $\Phi_K$ have been averaged.}
\label{fig:pxpz_el}
\end{center}
\end{figure}

\begin{figure}
\begin{center}
\resizebox{0.45\textwidth}{!}{\includegraphics{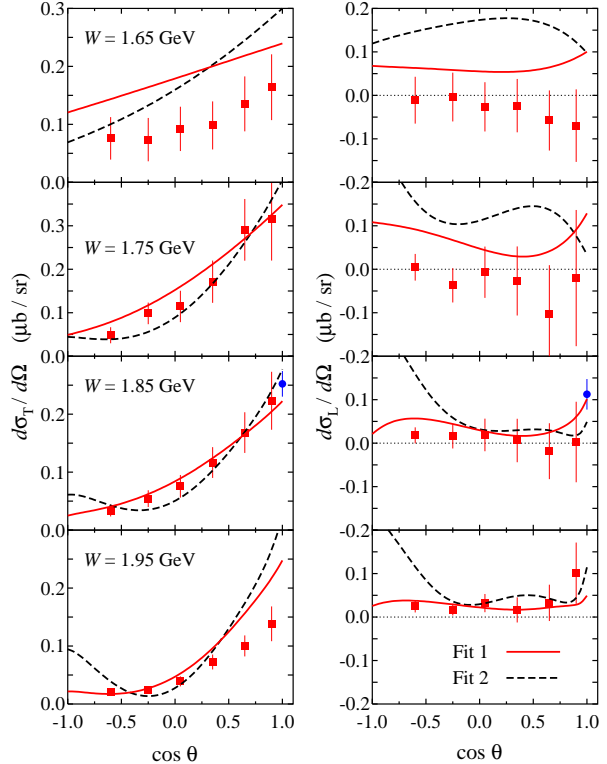}}
\caption{Comparison between the calculated
  longitudinal and transverse differential cross sections
  with the new {\small CLAS} data \cite{Ambrozewicz} at $Q^2=1$ GeV. 
  Except the two data points shown by solid circles at $W=1.85$ GeV
  \cite{Mohring:2002tr}, all data are not used in the fits.}
\label{fig:difcs_el}
\end{center}
\end{figure}

For the nucleon we use the standard Dirac and Pauli form factors, whereas
for the kaon we adopt the vector-meson-dominance model. The dependence of the
${\bar A}_{\ell\pm}^R$ multipole to the $Q^2$ is assumed to be
\begin{eqnarray}
  \label{eq:Q2_dependence}
  {\bar A}_{\ell\pm}^R(Q^2) &=& {\bar A}_{\ell\pm}^R(0)\, (1+a_1Q^2)\,
  e^{-a_2Q^2}~,
\end{eqnarray}
where $a_1$ and $a_2$ are fitting parameters.

In the fitting database we used 178 data points, including the
transfered $\Lambda$ polarization components $P'_{z'}$, $P'_{x'}$,
$P'_{z}$ and $P'_{x}$ data from {\small CLAS} \cite{Carman:2002se}. Note that
in the latter we do not use the data which are averaged over $Q^2$ 
and $d\Omega$ due to their accuracies and, furthermore, it is found
that they are relatively difficult to fit.
The comparison between experimental data and our calculation for the
transfered $\Lambda$ polarization components is shown in 
Fig.\,\ref{fig:pxpz_el}. Surprisingly, Fit 1 (obtained from 
fitting to the {\small SAPHIR} photoproduction data) yields a better 
explanation of the {\small CLAS} transfered $\Lambda$ polarization data.

Very recently, the {\small CLAS} collaboration has completed its analysis 
and published a relatively large number of kaon electroproduction
data \cite{Ambrozewicz}. These data are also not used in the fit,
for practical reasons, but we compare them with the result of the 
present analysis in Fig.\,\ref{fig:difcs_el}. As can be seen from this
figure, the same conclusion (as for the transfered $\Lambda$ polarization) 
can be drawn from this result.

\section{Conclusion}
We have analyzed the $K\Lambda$ photo- and electroproduction data by using
a multipole approach and the latest available experimental data. Our results 
show that the discrepancy between {\small CLAS} and {\small SAPHIR} photoproduction data
results in substantially different calculated electroproduction observables.
Surprisingly, the new {\small CLAS} electroproduction measurements can be better 
explained by a model that fits the {\small SAPHIR} data. Our next goal is to 
consider the $K\Sigma$ channels. These channels are of interest because they
can be related by using isospin symmetry \cite{Mart:1995wu}.

\section*{Acknowledgment}
This work has been partly supported by the Faculty of
Mathematics and Sciences, University of Indonesia, as well as 
the Hibah Pascasarjana grant.


\begin{thebibliography}{}
\bibitem{Mart:1999ed}
  T.~Mart and C.~Bennhold, 
  Phys.\ Rev.\ C {\bf 61} (1999) 012201.
\bibitem{Bradford:2005pt}
  R.~Bradford {\it et al.},
  Phys.\ Rev.\ C {\bf 73} (2006) 035202.
\bibitem{Glander:2003jw}
  K.~H.~Glander {\it et al.},
  Eur.\ Phys.\ J.\ A {\bf 19} (2004) 251.
\bibitem{Tiator:2003uu}
  L. Tiator {\it et al.}, 
  Eur.\ Phys.\ J.\ A {\bf 19} (2004) 55.
\bibitem{Mart:2006dk}
  T.~Mart and A.~Sulaksono,
  Phys.\ Rev.\ C {\bf 74} (2006) 055203.
\bibitem{Arndt:2003ga}
  R. A.~Arndt {\it et al.},
  Phys.\ Rev.\ C {\bf 69} (2004) 035208.
\bibitem{Mart:2004ug}
  T.~Mart, A.~Sulaksono and C.~Bennhold, 
  nucl-th/0411035.
\bibitem{Bartholomy:2004uz}
  O.~Bartholomy {\it et al.},
  Phys.\ Rev.\ Lett.\  {\bf 94} (2005) 012003.
\bibitem{Julia-Diaz:2006is}
  B.~Julia-Diaz, B.~Saghai, T.S.~Lee and F.~Tabakin, 
  Phys. Rev. C {\bf 73} (2006) 055204.
\bibitem{Mohring:2002tr}
  R.~M.~Mohring {\it et al.},
  Phys.\ Rev.\ C {\bf 67} (2003) 055205.
\bibitem{Carman:2002se}
  D.~S.~Carman {\it et al.},
  Phys.\ Rev.\ Lett.\  {\bf 90} (2003) 131804.
\bibitem{Ambrozewicz} 
  P. Ambrozewicz {\it et al.},
  hep-ex/0611036.
\bibitem{Mart:1995wu}
  T.~Mart, C.~Bennhold and C.~E.~Hyde-Wright,
  Phys.\ Rev.\ C {\bf 51} (1995) R1074.
\end{thebibliography}
\end{document}